\documentstyle[12pt]{article}
\def\int {\intop \limits}

\def\fnote#1{\footnote}
\begin{document}

\newcommand{\dst}[1]{\displaystyle{#1}}
\newcommand{\barl}{\begin{array}{rl}}
\newcommand{\ball}{\begin{array}{ll}}
\newcommand{\ear}{\end{array}}
\newcommand{\barc}{\begin{array}{c}}
\newcommand{\e}{\mbox{${\bf e}$}}
\newcommand{\J}{\mbox{${\bf J}$}}
\newcommand{\be}{\begin{equation}}
\newcommand{\ee}{\end{equation}}
\newcommand{\aq}[1]{\label{#1}}
\renewcommand \theequation{\thesection.\arabic{equation}}

\title{Coherent scattering of high-energy photon in a medium}
\author{V. N. Baier
and V. M. Katkov\\
Budker Institute of Nuclear Physics,\\ Novosibirsk, 630090,
Russia}

\maketitle

\begin{abstract}

The coherent scattering of photon in the Coulomb field
(the Delbr\"uck scattering) is considered for the momentum
transfer $\Delta \ll m$
in the frame of the quasiclassical operator method.
In high-energy region this process occurs over rather long distance.
The process amplitude is calculated taking into account the
multiple scattering of particles of the intermediate
electron-positron pair in a medium.
The result is the suppression of the process.
Limiting cases of weak and strong
effects of the multiple scattering are analyzed. The approach used is the
generalization of the method developed by authors for consideration
of the Landau-Pomeranchuk-Migdal effect.

\end{abstract}

\newpage
\section{Introduction}

The nonlinear effects of QED are due to the interaction of a photon with
electron-positron field. These processes are the photon-photon
scattering,  the coherent photon scattering (the elastic scattering
of a photon by the static Coulomb field called often the
Delbr\"uck scattering), the photon splitting into two photons, and
the coalescence of two photons into photon in the Coulomb field.
Among these processes the coherent scattering of photon
has been observed and investigated experimentally
(see reviews \cite{d2} and \cite{d3}) during last decades, the most
accurate measurement of the coherent scattering has been performed
not long ago in the Budker Institute of Nuclear Physics \cite{DS}.
The photon splitting into two photons in the Coulomb field
was observed for the first
time at Budker INP only recently \cite{PSp}.
Observation of photon-photon scattering is still a challenge.

History of the coherent photon scattering study can be found in the mentioned
reviews. There is a special interest to the process for heavy elements
because contributions of higher orders of $Z\alpha$ ($Z|e|$ is the charge
of nucleus, $e^2=\alpha=1/137,~\hbar=c=1 $) into the amplitude of
photon scattering are very important. This means that one needs the theory
which is exact with respect to the parameter $Z\alpha$.
The amplitudes of the coherent photon scattering valid for any $Z\alpha$
for high energy photon ($\omega \gg m$) and small scattering angle
(or small momentum transfer $\Delta$) were calculated in \cite{CW1},
\cite{CW2}. The approximate method of summing of the set of Feynman diagrams
with an arbitrary number of photons exchanged with the Coulomb source was
used. Another representation of these amplitudes (in the Coulomb field)
was found in \cite{MS1},
using the quasiclassical Green function of the Dirac equation
in the Coulomb field. This Green function in a spherically symmetrical
external field was obtained in \cite{LM1}
where the coherent photon scattering in the screened Coulomb potential
was investigated as well. Lately this Green function was calculated for
"localized potential", and the coherent photon scattering was analyzed
using it \cite{LM2}. Recently the process of the coherent photon scattering
was considered \cite{KS} in frame of the quasiclassical 
operator method (see e.g.
\cite{BKS}) which appears to be very adequate for consideration of this
problem.

The coherent photon scattering belong to the class of
electromagnetic processes
which in high-energy region occurs over rather long distance,
known as the formation length. Among other processes there are
the bremsstrahlung and the pair creation by a photon.
If anything happens to an electron
while traveling this distance, the emission can be disrupted.
Landau and Pomeranchuk~\cite{LP} showed that if the formation
length of bremsstrahlung becomes comparable to the distance over which
a mean angle of the multiple scattering becomes comparable with a
characteristic angle of radiation, the bremsstrahlung will be
suppressed. Migdal~\cite{M1} developed a quantitative
theory of this phenomenon which is known as the
Landau-Pomeranchuk-Migdal (LPM) effect.

A very successful series of
experiments (see \cite{E1}, \cite{E2}) was performed
at SLAC during last years. In these experiments the cross section
of bremsstrahlung of soft photons with energy from 200~KeV to
500~MeV from electrons with energy 8~GeV and 25~GeV is measured
with an accuracy of the order of a few percent. The LPM was
observed and investigated. These experiments were the
challenge for the theory since in all the previous papers calculations
are performed to logarithmic accuracy which is not enough for description
of the new experiment. The contribution of the Coulomb corrections (at least
for heavy elements) is larger then experimental errors and these corrections
should be taken into account.

Recently authors \cite{L1} developed the new
approach to the theory of the LPM effect
in frame of the quasiclassical operator method.
In it the cross section of bremsstrahlung process
in the photon energies region where the influence of
the LPM effect is very strong
was calculated with term $\propto 1/L$ , where $L$
is characteristic logarithm of the problem,
and with the Coulomb corrections
taken into account. In the photon energy region, where the LPM effect
is "turned off", the obtained cross section
gives the exact Bethe-Heitler cross section (within power accuracy) with
the Coulomb corrections. This important feature was absent in
the previous calculations. The contribution of an inelastic scattering of
a projectile on atomic electrons is also included.
We have analyzed (see \cite{L1}, \cite{L2})
the soft part of spectrum, including all the
accompanying effects: the boundary photon emission, the multiphoton radiation
and the influence of the polarization of a medium. Perfect agreement
of the theory and SLAC data was achieved in the whole interval of measured
photon energies.
Very recently we apply this approach to the process
of pair creation by a photon in \cite{L5}.

In the quasiclassical approximation the amplitude $M$
of the coherent photon scattering is described by diagram where
the electron-positron pair is created by the initial photon
with 4-momentum $k_1~(\omega,{\bf k}_1)$ and then annihilate into the final
photon with 4-momentum $k_2$. For high energy photon $\omega \gg m$
this process occurs over a rather long distance, known as
the time of life of the virtual state
\begin{equation}
l_f=\frac{\omega}{2q_c^2},
\label{1.1}
\end{equation}
where $q_c \geq m$ is the characteristic transverse momentum of the process,
the system $\hbar=c=1$ is used. When the virtual electron (or positron)
is moving in a medium it scatters on atoms. The mean square of
momentum transfer to the electron from a medium on the distance
$l_f$ is
\begin{equation}
q_s^2=4\pi Z^2\alpha^2n_aLl_f,\quad L(q_c)=\ln \frac{q_c^2}{q_{min}^2},\quad
q_{min}^2=a^{-2}+\Delta^2+\frac{m^4}{\omega^2},
\label{1.2}
\end{equation}
where $\alpha=e^2=1/137$, $Z$ is the charge of nucleus,
$n_a$ is the number density of atoms in the medium,
$\Delta$ is the photon momentum transfer ($\Delta=|{\bf k}_2-{\bf k}_1|
\ll q_c$),
$a$ is the screening radius of atom.

The coherent photon scattering amplitude $M$
can be obtained from general formulas for probabilities of
electromagnetic processes in the frame of 
the quasiclassical operator method (see e.g.
\cite{BKS}). It can be estimated as
\begin{equation}
M \sim \frac{\alpha \omega}{2\pi l_f n_a}\frac{q_s^2}{q_c^2}=
\frac{\alpha}{\pi n_a} q_s^2.
\label{1.3}
\end{equation}
We use the normalization
condition Im$M=\omega \sigma_p$ for the case $\Delta=0$,
where $\sigma_p$ is the total cross section of pair creation by a photon.

In the case of small momentum transfer $q_s \equiv \sqrt{q_s^2} \ll m$
we have in the region of small $\Delta \ll m$
\begin{equation}
q_c^2=m^2,\quad M \sim \frac{2Z^2 \alpha^3 \omega}{m^2}
\ln \frac{m^2}{a^{-2}+\Delta^2+\frac{m^4}{\omega^2}}.
\label{1.4}
\end{equation}

At an ultrahigh energy it is possible that $q_s \gg m$.
In this case the characteristic momentum transfer $q_c$ is
defined by the momentum transfer $q_{s}$. The self-consistency condition
is
\begin{equation}
q_c^2=q_s^2=\frac{2\pi \omega Z^2 \alpha^2 n_a L(q_c)}{q_c^2},
\label{1.5}
\end{equation}
where $L(q_c)$ is defined in (\ref{1.2}).
So using (\ref{1.3}) one gets for the estimate of the
coherent photon scattering amplitude $M$
(the influence of the multiple scattering manifests itself at the high 
photon energies such that $m^2 a/\omega \ll 1$)
\begin{equation}
M \sim \frac{2Z^2\alpha^3 \omega}{\Delta_s^2}
\sqrt{\ln \frac{\Delta_s^2}{a^{-2}+\Delta^2}},\quad
\Delta_s^2=\sqrt{2\pi \omega Z^2 \alpha^2 n_a} \gg \Delta^2.
\label{1.6}
\end{equation}

In the present paper we use consider the influence of the multiple scattering
on the process of the coherent photon scattering 
for $\Delta \ll q_c$. The theory
of the coherent photon scattering in the Coulomb field
in frame of the quasiclassical operator method is stated in Sec.2.
We give here an alternative presentation of approach developed in
\cite{KS} in a more formal way.
In Sec.3 we apply the method developed in \cite{L1}, \cite{L0}
to investigation of influence of the multiple scattering
on the process of the coherent photon scattering. The general formulas
describing this influence were derived and the asymptotic cases of
the strong and weak effects are analyzed.

\section{Coherent scattering of a photon \newline in the Coulomb field}
\setcounter{equation}{0}

\subsection{Formulation of approach}

The coherent scattering of a photon in the Coulomb field 
(the photon with momentum $k_{1}=(\omega_{1}, {\bf k}_{1})
\rightarrow k_{2}=(\omega_{2}, {\bf k}_{2})$) 
is represented by the electron
loop in the Coulomb field, and $\omega_1=\omega_2=\omega$. The 
corresponding amplitude is
\begin{equation}
T=-\frac{2\pi \alpha}{\omega}i\int_{}^{}d^4x_1 \int_{}^{}d^4x_2
Tr\left[\hat{e}_1\exp (ik_1x_1) G(x_1, x_2) \hat{e}_2^{\ast}
\exp (ik_2x_2) G(x_2, x_1)\right],
\label{3.1}
\end{equation}
where $\hat{e}=e_{\mu}\gamma^{\mu}$, $e_{\mu}$ is the photon polarization
vector, $G(x_1, x_2)$ is the electron Green function in the Coulomb
field the standard representation of which is
\begin{equation}
G(x_2, x_1)=\left\{\begin{array}{ll}
-&i\sum_{n}^{}\Psi_n^{(+)}(x_2)\overline{\Psi}_n^{(+)}(x_1),
\quad t_2 > t_1 \\
&i\sum_{n}^{}\Psi_n^{(-)}(x_2)\overline{\Psi}_n^{(-)}(x_1),
\quad t_2 < t_1
\end{array}\right\}.
\label{3.2}
\end{equation}
Here $\Psi_n^{(\pm)}(x_1)$ are the solution of Dirac equation in the Coulomb
field, signs $(+)$ and $(-)$ relate respectively to positive and negative
frequencies.

As well known, see e.g.\cite{BKF}, Sec.12,
in the noncovariant perturbation theory, which we use here,
in high-energy region ($\omega \gg m$)
the diagram contributes into the amplitude $T$ where
the electron-positron pair is first created by the initial photon
with 4-momentum $k_1$ and polarization $e_1$ in time $t_1$ and
then annihilate in time $t_2 > t_1$ into final photon
with 4-momentum $k_2$ and polarization $e_2$,
while the contribution of the interval in which $t_2 < t_1$
is of the order $m^2/\omega^2$. Such contribution
is neglected in the scope of our method.
Taking this into account and
substituting the Green function into the amplitude (\ref{3.1}) we find
\begin{equation}
T=\frac{2\pi \alpha}{\omega}i\sum_{n,m}^{}\int_{}^{}dt_1 \int_{}^{}dt_2
\vartheta(t_2-t_1) V_{nm}({\bf e}_1, {\bf k}_1, t_1)
V_{nm}^{\ast}({\bf e}_2, {\bf k}_2, t_2),
\label{3.3}
\end{equation}
where
\begin{equation}
V_{nm}({\bf e}, {\bf k}, t) = \int_{}^{}d^3 r\Psi_n^{(+)+}(x)
\mbox{\boldmath$\alpha$} {\bf e}\exp(-ikx) \Psi_m^{(-)}(x),
\label{3.4}
\end{equation}
It is evident that $V_{nm}({\bf e}, {\bf k}, t)$
(\ref{3.4}) is the matrix element of
pair creation by a photon in an external field. In the quasiclassical
operator method (see \cite{BKS}, Sec.3) the wave functions in an external
field can be presented in the form
\begin{equation}
\Psi_n^{(\pm)}({\bf r}, t) = <{\bf r}|\Psi^{(\pm)}({\bf P}, {\cal H})
\exp \left[\mp i({\cal H} \pm e\varphi)t \right] |n>,\quad
{\cal H}=\sqrt{m^2+{\bf P}^2},
\label{3.5}
\end{equation}
where $\varphi=\varphi({\bf r})$ is the potential of an atom,
$|n>$ is the state in the configuration space at the time $t=0$,
the functions $\Psi^{(\pm)}({\bf P}, {\cal H})$ have form of wave
functions for free particles in the momentum space
($\Psi^{(\pm)}({\bf p}, \varepsilon $). We substitute (\ref{3.5})
into (\ref{3.4}) and take into account completeness of states $|{\bf r}>$
($\displaystyle{\int_{}^{}|{\bf r}><{\bf r}| d^3r=I}$).
After this we convey consistently in
(\ref{3.4}) the operator $\exp (i{\bf k}{\bf r})$ to the right up to
$|m>$, and then the operator
$\exp \left[i({\cal H} + e\varphi)t \right]$ from
$<n|$ to $\exp \left[i({\cal H}({\bf k}-{\bf P}) - e\varphi)t \right]$.
As a result we
obtain the combination of operators
\begin{eqnarray}
&& V_{nm}({\bf e}, {\bf k}, t) = \Big<n|\Psi^{(+)+}({\bf P}(t))
\mbox{\boldmath$\alpha$} {\bf e}\Psi^{(-)}({\bf k}-{\bf P}(t))
\nonumber \\
&& \times \exp(i({\cal H}-e\varphi)t) \exp(i({\cal H}({\bf k}-{\bf P})
+e\varphi)t) \exp (i{\bf k}{\bf r})|m\Big>
\label{3.6a}
\end{eqnarray}
the disentanglement of which can be
performed in the standard way (see \cite{BKS}, Eqs.(3.7)-(3.11)):
\begin{eqnarray}
&& L_p(t)=\exp \left[i({\cal H} - e\varphi)t \right]
\exp \left[i({\cal H}({\bf k}-{\bf P}) + e\varphi)t \right]
\nonumber \\
&& \simeq \exp (i\omega t) {\rm T}\exp \left[
i\int_{0}^{t}\frac{kP(t)}{\omega-{\cal H}(t)}dt \right],\quad
P=({\cal H}, {\bf P}),
\label{3.6}
\end{eqnarray}
where ${\rm T}$ is the operator of the chronological product,
and the expression
\begin{equation}
R_p=\overline{\Psi}^{(+)}({\bf P}, {\cal H}) \hat{e}
\Psi^{(-)}({\bf k}-{\bf P}, {\cal H}({\bf k}-{\bf P}))
\label{3.7}
\end{equation}
becomes the Heisenberg operator depending on time ${\bf P}={\bf P}(t),~
{\cal H}={\cal H}(t)$. It can be expressed
in terms of two-component spinors. Substituting these results into
Eq.(\ref{3.4}) we find
\begin{eqnarray}
&& V_{nm}({\bf e}, {\bf k}, t)=\left<n|R_p(t){\rm T} \exp \left[
i\int_{0}^{t}\frac{kP(t)}{\omega-{\cal H}(t)}dt \right]
\exp (i{\bf k}{\bf r})|m\right>,
\nonumber \\
&& R_p(t)=i\varphi_{\overline{s}}^{+}\left(A(t)-
i\mbox{\boldmath$\sigma$}{\bf B}(t)\right)\varphi_s,\quad
A=\frac{1}{2 {\cal H} (\omega-{\cal H})}({\bf e}({\bf k}\times{\bf P})),
\nonumber \\
&& {\bf B} = \frac{1}{2 {\cal H} (\omega-{\cal H})}\left[
{\bf e} m\omega + ({\bf e}{\bf P}) ({\bf k}-2{\bf P}) \right],
\label{3.8}
\end{eqnarray}
where $\varphi_{\overline{s}}, \varphi_{s}$
are two-component spinors describing
polarization of created positron and electron.
Here the Coulomb gauge is used.

Substituting (\ref{3.8}) into (\ref{3.3})
and performing summation over states $|n>$ and $|m>$, as well as summation
over spin states $\overline{s}, s$
we have the amplitude
of the coherent scattering of a photon in the Coulomb field
\begin{eqnarray}
&&T=\frac{2\pi \alpha}{\omega}i\int_{}^{}dt_1 \int_{}^{}dt_2
\vartheta(t_2-t_1) S_{21}
\nonumber \\
&& S_{21}={\rm Tr}\left[m_2^+(t_2)
\exp[-i\mbox{\boldmath$\Delta$} {\bf r}] m_1(t_1)\right],\quad
\mbox{\boldmath$\Delta$}={\bf k}_2-{\bf k}_1,
\label{3.9}
\end{eqnarray}
where the operations Tr means sum of the diagonal matrix elements
both in the configuration and the spin spaces,
\begin{equation}
m_{1,2}(t)=m({\bf e}_{1,2}, k_{1,2}, t)=
\left(A(t)-i\mbox{\boldmath$\sigma$}{\bf B}(t)\right)
{\rm T}\exp \left[
i\int_{0}^{t}\frac{k_{1,2}P(t)}{\omega-{\cal H}(t)}dt \right],
\label{3.10}
\end{equation}
functions $A(t), {\bf B}(t)$ are defined in (\ref{3.8}).
Let us remind that ${\bf P}(t)$ is the momentum operator in the Heisenberg
picture:
\begin{equation}
{\bf P}(t)=\exp(-iHt){\bf P}\exp(iHt),\quad H= {\cal H}+V,
\label{3.11}
\end{equation}
where $V=e\varphi$

We present the evolution operator as
\begin{equation}
\exp(-iHt)=\exp(-i{\cal H}t)N(t),\quad
N(t)=\exp(i{\cal H}t)\exp(-i({\cal H}+V)t)
\label{3.12}
\end{equation}
Differentiating the last expression over time we find
\begin{equation}
\frac{dN(t)}{dt}=-i\exp(i{\cal H}t) V({\bf r})
\exp(-i({\cal H}+V)t) = -iV({\bf r}+{\bf v}t)N(t),\quad
{\bf v}=\frac{{\bf P}}{{\cal H}}
\label{3.13}
\end{equation}
The solution of this differential equation for the initial condition
$N(0)=1$ is
\begin{equation}
N(t)={\rm T}\exp \left[-i\int_{0}^{t}V({\bf r}+{\bf v}t')dt' \right],
\label{3.14}
\end{equation}

If the formation length of photon scattering is much longer than the
characteristic length of electron scattering
($\omega/m^2 \gg a$), one can present the dependence
on time of the operator ${\bf P}(t)$ in (\ref{3.8}) as
(see e.g. \cite{BKS}, Sec.7.1)
\begin{equation}
{\bf P}(t)=\vartheta(-t){\bf P}(-\infty)+\vartheta(t){\bf P}(\infty),\quad
{\bf P}(\pm \infty)=N^+(\pm \infty){\bf P}(t)N(\pm \infty).
\label{3.15}
\end{equation}
Let us introduce states convenient for further calculations
\begin{eqnarray}
&& |i>=N^+(-\infty)|{\bf p}_i>,\quad |f>=N^+(\infty)|{\bf p}_f>,
\nonumber \\
&& {\bf P}|{\bf p}_i>={\bf p}_i|{\bf p}_i>,\quad
{\bf P}|{\bf p}_f>={\bf p}_f|{\bf p}_f>.
\label{3.16}
\end{eqnarray}
The states $|i>$ and $|f>$ are the eigenvectors of the
operators ${\bf P}(-\infty)$ and ${\bf P}(\infty)$ correspondingly
\begin{eqnarray}
&& {\bf P}(-\infty)|i>=N^+(-\infty){\bf P} N(-\infty)N^+(-\infty)
|{\bf p}_i>={\bf p}_i |i>
\nonumber \\
&& {\bf P}(\infty)|f>=N^+(\infty){\bf P} N(\infty)N^+(\infty)
|{\bf p}_f>={\bf p}_f |f>.
\label{3.17}
\end{eqnarray}

The general expression for the amplitude of photon scattering
(Eqs.(\ref{3.9}) and (\ref{3.10})) is simplified significantly for
small momentum transfers $\Delta=|\mbox{\boldmath$\Delta$}|$
($\Delta \ll m,~\Delta/\varepsilon \ll 1/\gamma $).
With the accuracy $\sim \Delta/m$ we can put that the operator
${\bf P}$ commutes with $\exp[-i\mbox{\boldmath$\Delta$} {\bf r}]$ and
substitute the vector ${\bf k}_2$ in
the expression for $m_2(t_2)$ by the vector
${\bf k}_1$. Using (\ref{3.15}) and the states $|i>$ and $|f>$
(\ref{3.17}) for construction of the matrix element we have
\begin{eqnarray}
&& \left<f|m({\bf P}(t))\exp\left[-\frac{i}{2}\mbox{\boldmath$\Delta$}
{\bf r}\right]|i\right>=m_{fi}({\bf e}, {\bf p}(t))
\left<f|\exp\left[-\frac{i}{2}\mbox{\boldmath$\Delta$}
{\bf r}\right]|i\right>,
\nonumber \\
&&m_{fi}({\bf e}, {\bf p}(t))
=\left[\vartheta(-t)m({\bf e}, {\bf p}_i)+\vartheta(t)
m({\bf e}, {\bf p}_f) \right],
\label{3.18}
\end{eqnarray}
where we describe motion of particles of the created pair as a trajectory
in "the form of an angle".
Bearing in mind that the commutator
\begin{equation}
[r_i, v_j]=\frac{i}{{\cal H}}(\delta_{ij}-v_iv_j)
\label{3.19}
\end{equation}
we can discard operator T in the expression for operator $N(t)$ (\ref{3.14})
and with relativistic accuracy (i.e. with accuracy up to terms
$\sim m/\omega$) present the operator $N(t)$ as
\begin{equation}
N(t)=\exp \left[-i\int_{0}^{t}V(\mbox{\boldmath$\varrho$},z+t')dt' \right],
\label{3.20}
\end{equation}
where the axis $z$ is directed along the momentum of the initial photon
${\bf k}_1$, $\mbox{\boldmath$\varrho$}$ is two-dimensional vector transverse
to axis z.

Using (\ref{3.16}) and (\ref{3.20}) we calculate
the matrix element in (\ref{3.18})
\begin{eqnarray}
&& \left<f|\exp\left[-\frac{i}{2}\mbox{\boldmath$\Delta$}
{\bf r}\right]|i\right>=\int_{}^{} <f|{\bf r}><{\bf r}|i>
\exp\left[-\frac{i}{2}\mbox{\boldmath$\Delta$}
{\bf r}\right]d^3r
\nonumber \\
&& =\frac{i}{2\pi}f\left({\bf q}_{\perp}-
\frac{\mbox{\boldmath$\Delta$}_{\perp}}{2} \right)
\delta\left({\bf q}_{\parallel}-
\frac{\mbox{\boldmath$\Delta$}_{\parallel}}{2} \right),
\label{3.21}
\end{eqnarray}
where
\begin{eqnarray}
&& <f|{\bf r}>=<{\bf p}_f| N(\infty)|{\bf r}>=N(\infty)
<{\bf p}_f |{\bf r}>,\quad <{\bf p}_f |{\bf r}>=
\frac{\exp(-i{\bf p}_f{\bf r})}{(2\pi)^{3/2}},
\nonumber \\
&& <{\bf r}|i>=<{\bf r}| N^+(-\infty)|{\bf p}_i>=N^+(-\infty)
<{\bf r} |{\bf p}_i>,\quad <{\bf r}|{\bf p}_i>=
\frac{\exp(i{\bf p}_i{\bf r})}{(2\pi)^{3/2}},
\nonumber \\
&& f({\bf Q})=\frac{1}{2\pi i}\int_{}^{}
\exp [i{\bf Q}\mbox{\boldmath$\varrho$}+i\chi(\mbox{\boldmath$\varrho$})]
d^2\varrho,\quad
\chi(\mbox{\boldmath$\varrho$})=-\int_{-\infty}^{\infty}
V(\mbox{\boldmath$\varrho$},z)dz.
\label{3.22}
\end{eqnarray}
We introduced here vectors
\begin{equation}
{\bf p}_i \equiv {\bf p},\quad {\bf p}_f={\bf p}+{\bf q},
\label{3.23}
\end{equation}
the function $f({\bf Q})$ is the scattering amplitude in the eikonal
approximation.

Substituting Eq.(\ref{3.18}) into $T_{21}$ Eq.(\ref{3.9}) we
have
\begin{eqnarray}
&& S_{21}={\rm Tr}\left[m_2^+(t_2)
\exp[-i\mbox{\boldmath$\Delta$} {\bf r}] m_1(t_1)\right]
=\sum_{i,f}^{} {\rm Tr} \left[m_{fi}^{+}({\bf e}_2, {\bf p}(t_2))
m_{fi}({\bf e}_1, {\bf p}(t_1)) \right]
\nonumber \\
&&\times \left<i|\exp\left[\frac{i}{2}\mbox{\boldmath$\Delta$}
{\bf r}\right]|f\right> \left<f|\exp\left[-\frac{i}{2}
\mbox{\boldmath$\Delta$}{\bf r}\right]|i\right>
\nonumber \\
&& =\frac{\delta(\Delta_{\parallel})}{(2\pi)^2}
\int_{}^{}d^3p~ {\rm Tr} \left[m_{fi}^{+}({\bf e}_2, {\bf p}(t_2))
m_{fi}({\bf e}_1, {\bf p}(t_1)) \right]
\nonumber \\
&& \times \int_{}^{} d{\bf q}_{\perp}~
f\left({\bf q}_{\perp}-
\frac{\mbox{\boldmath$\Delta$}_{\perp}}{2} \right)
f^{\ast}\left({\bf q}_{\perp}+
\frac{\mbox{\boldmath$\Delta$}_{\perp}}{2} \right), \quad
q_{\parallel}=0.
\label{3.24}
\end{eqnarray}

\subsection{Helicity amplitudes}

In the above analysis we traced the transition to the expressions
calculated on the trajectory of particle "in the form of an angle"
in the momentum space. Precisely these trajectories determine
the amplitude of the coherent photon scattering for $\omega/m^2 \gg a$
(see Eq.(\ref{3.9})):
\begin{equation}
T_{21}=\frac{i\alpha}{(2\pi)^2}(2\pi \delta(\Delta_{\parallel}))
\int_{}^{} \frac{d^3p}{\omega} \int_{}^{}
d\sigma({\bf q}_{\perp}, \mbox{\boldmath$\Delta$})R_{21},
\label{3.25}
\end{equation}
where
\begin{eqnarray}
&& d\sigma({\bf q}_{\perp}, \mbox{\boldmath$\Delta$}) =
f\left({\bf q}_{\perp}-\frac{\mbox{\boldmath$\Delta$}_{\perp}}{2} \right)
f^{\ast}\left({\bf q}_{\perp}+
\frac{\mbox{\boldmath$\Delta$}_{\perp}}{2} \right) d{\bf q}_{\perp},
\nonumber \\
&& R_{21}= \int_{}^{}dt_1 \int_{}^{} dt_2 \vartheta(t_2-t_1)
{\rm Tr} \left[m_{fi}^{+}({\bf e}_2, {\bf p}(t_2))
m_{fi}({\bf e}_1, {\bf p}(t_1)) \right]
\nonumber \\
&& = \frac{1}{2}\int_{}^{}dt_1 \int_{}^{} dt_2 \vartheta(t_2-t_1)
{\cal L}({\bf e}_2, {\bf e}_1; {\bf v}_{2\perp}, {\bf v}_{1\perp})
\exp\left[-i\frac{\varepsilon}{\varepsilon'}\int_{t_1}^{t_2}
k_1v(t)dt \right],
\nonumber \\
&& {\cal L}({\bf e}_2, {\bf e}_1; {\bf v}_{2\perp}, {\bf v}_{1\perp})
= \frac{\omega^2}{\varepsilon'^2}\Bigg[({\bf e}_2^{\ast}{\bf n}
{\bf v}_{2\perp}) ({\bf e}_1{\bf n}{\bf v}_{1\perp})+
\frac{m^2}{\varepsilon^2}({\bf e}_2^{\ast}{\bf e}_1)
\nonumber \\
&&+\frac{(\omega-2\varepsilon)^2}{\omega^2}
({\bf e}_2^{\ast}{\bf v}_{2\perp}) ({\bf e}_1{\bf v}_{1\perp})\Bigg],
\quad {\bf v}_{1,2\perp}={\bf v}_{\perp}(t_{1,2})
\nonumber \\
&& v=\frac{p}{\varepsilon}=(1, {\bf v}),\quad \varepsilon'=\omega-
\varepsilon,\quad {\bf n}=\frac{{\bf k}_1}{\omega},\quad
({\bf v}_{1,2\perp}{\bf n})=0,
\label{3.26}
\end{eqnarray}
where $({\bf e}{\bf n}{\bf v}) \equiv ({\bf e}({\bf n}\times {\bf v}))$.

We calculate now $R_{21}$ in this expression for the trajectory
in "the form of an angle" (see (\ref{3.18})):
\begin{equation}
{\bf p}(t)={\bf p}~\vartheta(-t)+({\bf p}+{\bf q})\vartheta(t).
\label{3.27}
\end{equation}
We get for this case
\begin{eqnarray}
\hspace{-15mm}&& m^2 R_{21}=({\bf e}_2^{\ast}{\bf n}
{\bf g}) ({\bf e}_1{\bf n}{\bf g})
+\frac{(\omega-2\varepsilon)^2}{\omega^2}
({\bf e}_2^{\ast}{\bf g}) ({\bf e}_1{\bf g})+({\bf e}_2^{\ast}{\bf e}_1)
g_0^2;
\nonumber \\
\hspace{-15mm}&& {\bf g} = \frac{{\bf p}_{\perp}+{\bf q}_{\perp}}
{m^2+({\bf p}_{\perp}+{\bf q}_{\perp})^2} -
\frac{{\bf p}_{\perp}}
{m^2+{\bf p}_{\perp}^2},\quad g_0=
\frac{m}{m^2+({\bf p}_{\perp}+{\bf q}_{\perp})^2}
-\frac{m}{m^2+{\bf p}_{\perp}^2}.
\label{3.28}
\end{eqnarray}
The functions ${\bf g}$ and $g_0$ have very important properties:
\begin{equation}
{\bf g}({\bf q}_{\perp}=0) = 0,\quad g_0({\bf q}_{\perp}=0) = 0.
\label{3.29}
\end{equation}

In the case of complete screening when the screening
radius $a \ll \omega/m^2$ the amplitude of photon
scattering $T_{21}$ (\ref{3.25}) is imaginary at any
$\mbox{\boldmath$\Delta$}$. Taking into account that at
$\mbox{\boldmath$\Delta$}=0$ this amplitude is connected due to
unitary relation with the known probability of pair photoproduction,
see e.g. \cite{BKF}, \cite{BLP}, we will calculate the difference
\[
\delta R_{21} = T_{21}(\mbox{\boldmath$\Delta$}) - T_{21}(0).
\]
The interval $|{\bf q}_{\perp}| \leq |\mbox{\boldmath$\Delta$}| \ll m$
contributes into this difference and one can expand the functions
${\bf g}, g_0$ as a power series in ${\bf q}_{\perp}$:
\begin{equation}
{\bf g} \simeq \frac{{\bf q}_{\perp}}
{m^2+{\bf p}_{\perp}^2} -
\frac{2{\bf p}_{\perp}({\bf q}_{\perp}{\bf p}_{\perp})}
{(m^2+{\bf p}_{\perp}^2)^2},\quad g_0 \simeq -
\frac{2m({\bf q}_{\perp}{\bf p}_{\perp})}
{(m^2+{\bf p}_{\perp}^2)^2}.
\label{3.30}
\end{equation}
We introduce the dimensionless variable ${\bf u}={\bf p}_{\perp}/m$
and carry on averaging over angles of ${\bf u}$ (the azimuthal angle
of the component of electron momentum in the plane which is perpendicular
to the direction of initial photon ${\bf n}$) using formulas
\begin{equation}
\overline{u_i u_j} = \frac{u^2}{2}\delta_{ij},\quad
\overline{u_i u_j u_k u_l} = \frac{u^4}{8}\left(\delta_{ij}\delta_{kl}+
\delta_{ik}\delta_{jl}+\delta_{il}\delta_{jk} \right),
\label{3.31}
\end{equation}
where $\delta_{ij}$ is the two-dimensional Kronecker delta.
Substituting Eq.(\ref{3.30}) into (\ref{3.28}) and using (\ref{3.31})
we get
\begin{eqnarray}
&&  m^4 \zeta^4 R_{21}=({\bf e}_2^{\ast}{\bf n}
{\bf q}_{\perp}) ({\bf e}_1{\bf n}{\bf q}_{\perp})
+\frac{(\varepsilon'-\varepsilon)^2}{\omega^2}
({\bf e}_2^{\ast}{\bf q}_{\perp}) ({\bf e}_1{\bf q}_{\perp})
\nonumber \\
&&+{\bf q}_{\perp}^2({\bf e}_2^{\ast}{\bf e}_1)
\left[\left(1-\frac{2\varepsilon \varepsilon'}{\omega^2}\right)(\zeta^2-1)+
\frac{4\varepsilon \varepsilon'}{\omega^2}(\zeta-1) \right],\quad
\zeta=1+{\bf u}^2.
\label{3.32}
\end{eqnarray}

It is convenient to describe the process of photon scattering in terms
of helicity amplitudes. We choose the polarization vectors with helicity
$\lambda$
\begin{eqnarray}
&&{\bf e}_{\lambda}=\frac{1}{\sqrt{2}}\left({\bf e}_1
+i\lambda{\bf e}_2 \right),\quad
{\bf e}_1= \mbox{\boldmath$\nu$}=\frac{\mbox{\boldmath$\Delta$}}
{|\mbox{\boldmath$\Delta$}|},\quad
{\bf e}_2={\bf n}\times \mbox{\boldmath$\nu$},\quad \lambda=\pm 1,
\nonumber \\
&& {\bf e}_{\lambda}{\bf e}_{\lambda}^{\ast}=1,\quad
{\bf e}_{\lambda} {\bf e}_{-\lambda}^{\ast}=0,\quad
{\bf e}_{\lambda}\times {\bf n}=i\lambda{\bf e}_{\lambda},
\label{3.33}
\end{eqnarray}
where ${\bf n}$ is defined in Eq.(\ref{3.26}).
There are two independent helicity amplitudes:
\[
M_{++}=M_{--},\quad M_{+-}=M_{-+},
\]
where the first subscript is the helicity of the initial photon and
the second is the helicity of the final photon.
When the initial photons are unpolarized the differential cross section
of scattering summed over final photons polarization contains the
combination
\begin{equation}
2[|M_{++}|^2+|M_{+-}|^2].
\label{3.34}
\end{equation}
Substituting (\ref{3.33}) into (\ref{3.32}) we find
\begin{eqnarray}
&& R_{++}=\frac{{\bf q}_{\perp}^2}{m^4 \zeta^4}
\left[\left(1-\frac{2\varepsilon \varepsilon'}{\omega^2}\right)(\zeta^2-1)+
\frac{4\varepsilon \varepsilon'}{\omega^2}(\zeta-1) \right],
\nonumber \\
&& R_{+-}=-\frac{2{\bf q}_{\perp}^2}{m^4 \zeta^4}
\frac{\varepsilon \varepsilon'}{\omega^2} \cos 2\varphi,\quad
\cos \varphi = \frac{{\bf q}\mbox{\boldmath$\nu$}}{q}.
\label{3.35}
\end{eqnarray}

We define the amplitude of the coherent photon scattering $M_{21}$ as
\begin{equation}
T_{21}=\frac{\pi}{\omega}\delta(\Delta_{\parallel}) M_{21},
\label{3.36}
\end{equation}
then the cross section of the coherent photon scattering has a form
\begin{equation}
d\sigma_s=\frac{1}{2\pi \delta(\Delta_{\parallel})} |T_{21}|^2
\frac{d^3k_2}{(2\pi)^3}=\frac{1}{16\pi^2}|M_{21}|^2 d\Omega_2.
\label{3.37}
\end{equation}

Substituting (\ref{3.36}), (\ref{3.32}) and (\ref{3.35}) into Eq.(\ref{3.25}) we
obtain for the helicity amplitudes
\begin{equation}
M_{\lambda \lambda'}= \frac{i\alpha}{2\pi}m^2
\int_{0}^{\omega} d\varepsilon \int_{1}^{\infty}d\zeta
\int_{}^{}
d\sigma({\bf q}_{\perp}, \mbox{\boldmath$\Delta$})R_{\lambda \lambda'},
\label{3.38}
\end{equation}
where $\lambda \lambda'$ is $++$ or $+-$.
For the chosen normalization of the helicity amplitudes
the unitary relation with total cross section of pair photoproduction
$\sigma_p$ is
\begin{equation}
\frac{1}{\omega}{\rm Im}~M_{++}=\sigma(\gamma \rightarrow e^+e^-)
\equiv \sigma_p.
\label{3.39}
\end{equation}
So we present the helicity amplitude $M_{++}$ as
\begin{eqnarray}
&& M_{++}=\delta M_{++} + i \omega \sigma_p,
\nonumber \\
&&\delta M_{++} = \frac{i\alpha}{2\pi}m^2
\int_{0}^{\omega} d\varepsilon \int_{1}^{\infty}d\zeta
\int_{}^{} (d\sigma({\bf q}_{\perp}, \mbox{\boldmath$\Delta$})
-d\sigma({\bf q}_{\perp}, 0))R_{++}.
\label{3.40}
\end{eqnarray}
Since the forward scattering amplitude ($\Delta=0$) with helicity flip
vanishes we have for amplitude $M_{+-}$
\begin{equation}
M_{+-}=\delta M_{+-} = \frac{i\alpha}{2\pi}m^2
\int_{0}^{\omega} d\varepsilon \int_{1}^{\infty}d\zeta
\int_{}^{}
d\sigma({\bf q}_{\perp}, \mbox{\boldmath$\Delta$})R_{+-}.
\label{3.41}
\end{equation}
Integrals over ${\bf q}_{\perp}$ in Eqs.(\ref{3.40}) and (\ref{3.41})
we denote $\overline{\delta {\bf q}_{\perp}^2}$ and
$\overline{{\bf q}_{\perp}^2\cos 2\varphi}$. We calculate these integrals
using the scattering amplitude $f({\bf Q})$
in the eikonal approximation Eq.(\ref{3.22}). We find
\begin{eqnarray}
&& \overline{\delta {\bf q}^2} \equiv Z_1 =
\int_{}^{}  {\bf q}_{\perp}^2
(d\sigma({\bf q}_{\perp}, \mbox{\boldmath$\Delta$})
-d\sigma({\bf q}_{\perp}, 0))
\nonumber \\
&& = \int_{}^{} d{\bf q}_{\perp} {\bf q}_{\perp}^2
\left[f^{\ast}\left({\bf q}_{\perp}+
\frac{\mbox{\boldmath$\Delta$}_{\perp}}{2} \right)
f\left({\bf q}_{\perp}-
\frac{\mbox{\boldmath$\Delta$}_{\perp}}{2} \right)
- f^{\ast}\left({\bf q}_{\perp}\right)f\left({\bf q}_{\perp}\right)
\right]
\nonumber \\
&& = 2\pi \int_{0}^{\infty} (\chi'(\varrho))^2
(J_0(\Delta \varrho)-1)\varrho d\varrho,
\nonumber \\
&& -\overline{{\bf q}^2\cos 2\varphi} \equiv Z_2 =
2\pi \int_{0}^{\infty} (\chi'(\varrho))^2
J_2(\Delta \varrho)\varrho d\varrho.
\label{3.42}
\end{eqnarray}
For the screened Coulomb potential
\begin{eqnarray}
&& U(r)=\frac{Z\alpha}{r}\exp(-r/a),\quad \chi'(\varrho)=\frac{2Z\alpha}{a}
K_1\left(\frac{\varrho}{a} \right),
\nonumber \\
&& Z_1=-4\pi Z^2\alpha^2 F_2\left(\frac{\Delta a}{2} \right),\quad
F_2(z)=\frac{2z^2+1}{z\sqrt{1+z^2}}\ln \left(z+\sqrt{1+z^2} \right)-1,
\nonumber \\
&& Z_2=4\pi Z^2\alpha^2 F_1\left(\frac{\Delta a}{2} \right),\quad
F_1(z)=1-\frac{1}{z\sqrt{1+z^2}}\ln \left (z+\sqrt{1+z^2}\right).
\label{3.43}
\end{eqnarray}
The functions $F_1(x)$ and $F_2(x)$ encounter in the radiation theory
and in the theory of Landau-Pomeranchuk-Migdal (LPM) effect.
In this derivation the following relations have been used
\begin{eqnarray}
&& \int_{0}^{\infty}xK_1^2(x)J_2(\beta x)
=-\frac{1}{\beta}\int_{0}^{\infty}(xK_1(x))^2\frac{d}{dx}
\left(\frac{J_1(\beta x)}{x} \right)
\nonumber \\
&&=\frac{1}{2}-\frac{2}{\beta}\int_{0}^{\infty}K_0(x)K_1(x)xJ_1(\beta x)
\nonumber \\
&& =\frac{1}{2}+\frac{1}{\beta}\int_{0}^{\infty}
\frac{d}{dx}(K_0(x))^2xJ_1(\beta x)
=\frac{1}{2}-\int_{0}^{\infty}(K_0(x))^2J_0(\beta x),
\nonumber \\
&&\int_{0}^{\infty}x K_0^2(x) xdx=\frac{1}{2}.
\label{4.10}
\end{eqnarray}

The cross section of $e^-e^+$ pair photoproduction in the case of
complete screening ($a \ll \omega/m^2$) to within terms $\sim m/\omega$
has a form (see e.g. Eq.(19.17) in \cite{BKF})
\begin{equation}
\sigma_p=\frac{28Z^2\alpha^3}{9m^2}\left[\ln (ma)+\frac{1}{2}-f(Z\alpha)
-\frac{1}{42} \right],
\label{3.44}
\end{equation}
where
\begin{equation}
f(\xi)={\rm Re}\left[\psi(1+i\xi)-\psi(1)\right]
=\xi^2\sum_{n=1}^{\infty}\frac{1}{n(n^2+\xi^2)},
\label{3.45}
\end{equation}
here $\psi(x)$ is the logarithmic derivative of the gamma function.
Substituting (\ref{3.42})-(\ref{3.44}) into Eqs.(\ref{3.40}) and (\ref{3.41})
we obtain
\begin{equation}
{\rm Im}~M_{\lambda \lambda'}=\frac{4Z^2\alpha^3 \omega}{m^2}
\int_{0}^{1}dx\int_{0}^{1}dy~ \mu_{\lambda \lambda'}f_{\lambda \lambda'},
\label{3.46}
\end{equation}
where
\begin{eqnarray}
&& \mu_{++}=1-2x(1-x)+4x(1-x)y(1-y),\quad \mu_{+-}=x(1-x)y^2;
\nonumber \\
&& f_{++}=-\frac{1}{2}F_2\left(\frac{\Delta a}{2}\right)+
\ln (ma)+\frac{1}{2}-f(Z\alpha)-\frac{1}{42}
\nonumber \\
&&=\ln (ma)-\frac{2s^2+1}{2s\sqrt{1+s^2}}\ln \left(s+\sqrt{1+s^2} \right)
-f(Z\alpha)+\frac{41}{42},
\nonumber \\
&&f_{+-}=F_1\left(s\right),\quad s=\frac{\Delta a}{2}.
\label{3.47}
\end{eqnarray}
Here we passed to the variables $x=\varepsilon/\omega,~y=1/\zeta$.

\subsection{The coherent photon scattering in different cases}

The important property of Eq.(\ref{3.46}) is that the dependence
on the screening radius $a$ originates in it from the Born approximation.
In this approximation in the case of arbitrary screening
the radius $a$ enters only in the combination
\begin{equation}
\frac{1}{a^2}+q_{\parallel}^2,\quad q_{\parallel}
=\frac{m^2\omega}{2\varepsilon \varepsilon'}(1+u^2)=\frac{q_{m}}{x(1-x)y},
\quad q_m=\frac{m^2}{2\omega}.
\label{3.48}
\end{equation}
Since in the course of derivation of Eq.(\ref{3.46}) we did not integrate
by parts over the variables $x$ and $y$, we can extend Eq.(\ref{3.47}) on the
case of arbitrary screening making the substitution
\begin{equation}
\frac{1}{a} \rightarrow \sqrt{q_{\parallel}^2+a^{-2}} \equiv
q_{ef},\quad s=\frac{\Delta}{2q_{ef}}
\label{3.49}
\end{equation}
Recently the amplitudes of the coherent photon scattering were derived in
\cite{KS} for arbitrary interrelation between parameters in the
Moli\`ere potential.

We calculate the real part of scattering amplitude using the
dispersion relations method. We can apply this method directly
since in the region of momentum transfer under consideration
($q_{\perp} \ll m$) the dependence of amplitude
$T_{21} \propto M_{21}/\omega$ on $q_{\parallel}$ has the form
(see Eqs.(\ref{3.9})-(\ref{3.14}), (\ref{3.20}), (\ref{3.23}))
\begin{equation}
T(q_{\parallel})=\int_{}^{}dt_1\int_{}^{}dt_2 F(t_1, t_2)
\vartheta(t_2-t_1)\exp [-iq_{\parallel}(t_2-t_1)].
\label{3.50}
\end{equation}
Thus the real and imaginary part of the amplitude $M_{21}/\omega$
are connected by the Sokhotsky-Plemelj formulas
\begin{equation}
{\rm Re}~f(z)=\frac{{\cal P}}{\pi}\int_{-\infty}^{\infty}
\frac{{\rm Im}~ f(z')}{z'-z} dz'
\label{3.51}
\end{equation}
Using this transformation and Eqs.(\ref{3.46}), (\ref{3.47})
we obtain
\begin{equation}
{\rm Re}~M_{\lambda \lambda'}=\frac{4Z^2\alpha^3 \omega}{m^2}
\int_{0}^{1}dx\int_{0}^{1}dy~ \mu_{\lambda \lambda'}
\varphi_{\lambda \lambda'},
\label{3.52}
\end{equation}
where
\begin{eqnarray}
&& \varphi_{\lambda \lambda'}=\frac{1}{\pi}\int_{0}^{\infty}
\left[f_{\lambda \lambda'}(z+q_{\parallel})-
f_{\lambda \lambda'}(|z-q_{\parallel}|) \right]\frac{dz}{z}
\nonumber \\
&& =\frac{1}{\pi}\int_{0}^{\infty}
\left[f_{\lambda \lambda'}(q_{\parallel}(z+1))-
f_{\lambda \lambda'}(q_{\parallel}|z-1|) \right]\frac{dz}{z}
\label{3.53}
\end{eqnarray}

We consider now different limiting cases depending on interrelation
between $a$ and $\omega/m^2$. In the case of complete screening
($a \ll \omega/m^2$) we can use directly
Eqs.(\ref{3.46}), (\ref{3.47}). In this case the functions
$f_{\lambda \lambda'}$ are independent of $x$ and $y$ and the corresponding
integrals are
\begin{equation}
\int_{0}^{1}dx\int_{0}^{1}dy~\mu_{++}=\frac{7}{9},\quad
\int_{0}^{1}dx\int_{0}^{1}dy~\mu_{+-}=\frac{1}{18}.
\label{3.54}
\end{equation}
For the scattering amplitudes we have
\begin{eqnarray}
&& {\rm Im}~M_{++}=\frac{28Z^2\alpha^3\omega}{9m^2}f_{++},\quad
{\rm Im}~M_{+-}=\frac{2Z^2\alpha^3\omega}{9m^2}f_{+-},
\nonumber \\
&&{\rm Re}~M_{++}=0,\quad {\rm Re}~M_{+-}=0
\label{3.55}
\end{eqnarray}
The results obtained are consistent with Eqs.(33) and (36)
of \cite{LM1} where calculation has been done for the Moli\`ere potential.

In the case $a \gg \omega/m^2$ (the screening radius is very large, or
in other words we consider the photon scattering in the Coulomb field)
we have, taking into account Eqs.(\ref{3.47}), (\ref{3.49})
\begin{eqnarray}
\hspace{-15mm}&& f_{++}= \ln \frac{m}{q_{\parallel}}-
\frac{2s_c^2+1}{s_c\sqrt{1+s_c^2}}
\ln \left(s_c+\sqrt{1+s_c^2} \right)
-f(Z\alpha)+\frac{41}{42},
\nonumber \\
\hspace{-15mm}&&f_{+-}=1-\frac{1}{s_c\sqrt{1+s_c^2}}
\ln \left(s_c+\sqrt{1+s_c^2} \right),\quad s_c=\frac{\Delta}{2q_{\parallel}}
=\frac{\Delta \omega}{m^2}x(1-x)y.
\label{3.56}
\end{eqnarray}
The expressions for amplitudes $M_{++}$ and $M_{+-}$ in this case
are consistent with the results obtained in \cite{CW1}.
At $\Delta=0$ we find known result (see Eq.(8.1) in \cite{CW1})
\begin{eqnarray}
&& f_{++}= \ln \left[\frac{2\omega}{m}x(1-x)y\right]-\frac{1}{2}
-f(Z\alpha)+\frac{41}{42},
\nonumber \\
&& \varphi_{++}=\frac{1}{\pi}\int_{0}^{\infty}
\ln \frac{z+1}{|z-1|}\frac{dz}{z}
=\frac{2}{\pi}\int_{0}^{1}\ln \frac{1+z}{1-z}\frac{dz}{z}
=\frac{\pi}{2},
\nonumber \\
&& M_{++}=i\frac{28Z^2\alpha^3\omega}{9m^2}\left[\ln \frac{2\omega}{m}
-f(Z\alpha)-\frac{109}{42}-i\frac{\pi}{2} \right],\quad M_{+-}=0.
\label{3.57}
\end{eqnarray}
At $\Delta=0$ and arbitrary interrelation between $a$ and $\omega/m^2$
we get
\begin{eqnarray}
\hspace{-10mm}&& f_{++}= \ln \frac{m}{q_{ef}}-\frac{1}{2}
-f(Z\alpha)+\frac{41}{42},
\nonumber \\
\hspace{-10mm}&& \varphi_{++}=\frac{1}{2\pi}\int_{0}^{\infty}
\ln \frac{\delta^2(z+1)^2+x^2(1-x)^2y^2}
{\delta^2(z-1)^2+x^2(1-x)^2y^2}\frac{dz}{z},
\quad \delta=\frac{am^2}{2\omega}=\frac{m}{2\omega}
\frac{a}{\lambda_c}.
\label{3.58}
\end{eqnarray}
The photon scattering amplitude in this case for arbitrary value of
parameter $\delta$ was found recently in \cite{LM2}.

\section{Influence of the multiple scattering on the coherent scattering
of a photon}
\setcounter{equation}{0}

\subsection{Basic equations}

When a photon is propagating in a medium it
dissociates with probability $\propto \alpha$ into an electron-positron
pair. The virtual electron and positron interact with a medium
and can scatter on atoms. In this scattering 
the electron and positron interaction with the Coulomb field
in the course of the coherent scattering of photon is involved also.
There is a direct analogue with the Landau-Pomeranchuk-Migdal (LPM)
effect: the influence of the multiple scattering on process of the
bremsstrahlung and pair creation by a photon in a medium at high energy
\cite{LP}, \cite{M1}. However there is the difference: in the LPM effect
the particles of electron-positron pair created by a
photon are on the mass shell while in the process
of the coherent scattering of photon this particles are off the mass shell,
but in the high energy region (this is the only region where the influence
of the multiple scattering is pronounced) the shift from the mass shell
is relatively small.
To include this scattering into consideration the amplitude
$R_{21}$ Eq.(\ref{3.25})
should be averaged over all possible trajectories of electron and positron
with the weight function $d\sigma({\bf q}, \mbox{\boldmath$\Delta$})$.
This operation can be performed with the aid of the distribution function
averaged over the atomic positions of scatterers in the medium.
This procedure was worked out in details in \cite{L0} (Sec.2),
\cite{L1} (Sec.2). The amplitude of the coherent photon scattering for
a photon propagating in a medium can be derived in the same way as
Eqs.(2.4)-(2.6) of \cite{L0}:
\begin{eqnarray}
&& M({\bf e}_1, {\bf e}_2, \mbox{\boldmath$\Delta$})
=\frac{i\alpha}{(2\pi)^2n_a}\int_{}^{}d^3p
\exp\left(-i\frac{\varepsilon}{\varepsilon'}\tau \right)
\int_{}^{}d{\bf v}'\int_{}^{}d{\bf r}'
{\cal L}\left({\bf e}_1,{\bf e}_2;\mbox{\boldmath$\vartheta$}',
\mbox{\boldmath$\vartheta$} \right)
\nonumber \\
&& \times F\left({\bf r}',{\bf v}',\tau;{\bf r}, {\bf v} \right)
\exp\left[i\frac{\varepsilon}{\varepsilon}'{\bf k}_1
({\bf r}'-{\bf r}) \right].
\label{4.0}
\end{eqnarray}
The distribution function
$F\left({\bf r}',{\bf v}',\tau;{\bf r}, {\bf v} \right)$ satisfies the
kinetic equation
\begin{eqnarray}
&& \frac{\partial F\left({\bf r}',{\bf v}',\tau \right)}{\partial \tau}
+{\bf v}'\frac{\partial F\left({\bf r}',{\bf v}',\tau \right)}
{\partial {\bf r}'}
\nonumber \\
&& =n_a\int_{}^{}d\sigma({\bf v}', {\bf v}'', \mbox{\boldmath$\Delta$})
\left[F\left({\bf r}',{\bf v}'',\tau \right)-
F\left({\bf r}',{\bf v}',\tau \right) \right],
\label{4.0a}
\end{eqnarray}
and the initial condition
\[
F\left({\bf r}',{\bf v}', 0;{\bf r}, {\bf v} \right)=
\delta({\bf r}-{\bf r}')\delta({\bf v}-{\bf v}'),
\]
here $n_a$ is the number density of atoms in a medium and in the case
of the screened Coulomb potential
\begin{equation}
d\sigma({\bf v}', {\bf v}'', \mbox{\boldmath$\Delta$})=
d\sigma({\bf q}, \mbox{\boldmath$\Delta$}),\quad
{\bf q}=\varepsilon({\bf v}'-{\bf v}).
\label{4.0b}
\end{equation}
The combinations entering in Eq.(\ref{3.26}) are
\begin{eqnarray}
&& \int_{t_1}^{t_2} kv(t)dt=\omega \tau -
{\bf k}_1({\bf r}(t_2)-{\bf r}(t_1)) \rightarrow
\omega \tau - {\bf k}_1({\bf r}'-{\bf r}),
\nonumber \\
&& {\cal L}\left({\bf e}_1,{\bf e}_2;{\bf v}_2, {\bf v}_1 \right)
\rightarrow {\cal L}\left({\bf e}_1,{\bf e}_2;\mbox{\boldmath$\vartheta$}',
\mbox{\boldmath$\vartheta$} \right)
\label{4.0c}
\end{eqnarray}

We can proceed with further calculation (integration over
${\bf v}',~{\bf r}'$) of (\ref{4.0})
by analogy with the procedure used in Eqs.(2.7)-(2.18) of \cite{L0}.
As a result we obtain for the differential cross section
of the coherent photon scattering with the multiple 
scattering taken into account
\begin{equation}
\frac{d\sigma}{d\Omega}=\frac{1}{16\pi^2}
\left|M({\bf e}_1, {\bf e}_2, \mbox{\boldmath$\Delta$})\right|^2,
\label{4.1}
\end{equation}
where
\begin{eqnarray}
&& M({\bf e}_1, {\bf e}_2, \mbox{\boldmath$\Delta$})
=\frac{2i\alpha m^2\omega}{n_a} \int_{}^{}
\frac{d\varepsilon}{\varepsilon \varepsilon'}
\int_{o}^{\infty}dt \exp(-it)\Bigg[({\bf e}_2^{\ast}{\bf e}_1)
\varphi_0(0, t)
\nonumber \\
&&-i ({\bf e}_1{\bf n}\mbox{\boldmath$\nabla$})
({\bf e}_2^{\ast}{\bf n} \mbox{\boldmath$\varphi$}(0,t))
-i\frac{(\omega-2\varepsilon)^2}{\omega^2}
({\bf e}_1\mbox{\boldmath$\nabla$})
({\bf e}_2^{\ast}\mbox{\boldmath$\varphi$}(0,t))\Bigg].
\label{4.2}
\end{eqnarray}
In derivation we have changed variables into $t, \mbox{\boldmath$\varrho$}$
as in Eq.(2.7) of \cite{L1}.
The function $\varphi_{\mu}(\mbox{\boldmath$\varrho$}, t),~
\varphi_{\mu}=(\varphi_0, \mbox{\boldmath$\varphi$})$ satisfies the
equation
\begin{eqnarray}
\hspace{-10mm}&& i\frac{\partial \varphi_{\mu}}{\partial t}
={\cal H}\varphi_{\mu},\quad
{\cal H}={\bf p}^2-iV(\mbox{\boldmath$\varrho$}, \mbox{\boldmath$\Delta$}),
\quad {\bf p}=-i\mbox{\boldmath$\nabla$}_{\mbox{\boldmath$\varrho$}},
\nonumber \\
\hspace{-10mm}&& V(\mbox{\boldmath$\varrho$}, \mbox{\boldmath$\Delta$})
=\frac{2\varepsilon \varepsilon' n_a}{\omega m^4}
\int_{}^{}\left(1-\exp(i{\bf q} \mbox{\boldmath$\varrho$}) \right)
f\left({\bf q}_{\perp}-\frac{\mbox{\boldmath$\Delta$}_{\perp}}{2m} \right)
f^{\ast}\left({\bf q}_{\perp}+
\frac{\mbox{\boldmath$\Delta$}_{\perp}}{2m} \right)d^2q,
\label{4.3}
\end{eqnarray}
with the initial conditions
\begin{equation}
\varphi_{0}(\mbox{\boldmath$\varrho$}, 0)
=\delta(\mbox{\boldmath$\varrho$}),\quad
\mbox{\boldmath$\varphi$}(\mbox{\boldmath$\varrho$}, 0)
={\bf p} \delta(\mbox{\boldmath$\varrho$}).
\label{4.4}
\end{equation}

The potential $V(\mbox{\boldmath$\varrho$}, 0) \equiv
V(\mbox{\boldmath$\varrho$})$ was used in the theory of the LPM
effect (see \cite{L1} (Sec.2)):
\begin{eqnarray}
&& V(\mbox{\boldmath$\varrho$})=Q\mbox{\boldmath$\varrho$}^2
\left(L_1+\ln \frac{4}{\mbox{\boldmath$\varrho$}^2}-2C \right), \quad
Q=\frac{2\pi Z^2\alpha^2\varepsilon \varepsilon'n_a}{m^4\omega},\quad
L_1=\ln \frac{a_{s2}^2}{\lambda_c^2},
\nonumber \\
&& \frac{a_{s2}}{\lambda_c}=183Z^{-1/3}{\rm e}^{-f},\quad
f=f(Z\alpha)=(Z\alpha)^2\sum_{k=1}^{\infty}\frac{1}{k(k^2+(Z\alpha)^2)},
\label{4.5}
\end{eqnarray}
where $C=0.577216 \ldots$ is Euler's constant. We can restrict ourselves to
calculation of the difference of the potentials only:
\begin{eqnarray}
&& \Delta V(\mbox{\boldmath$\varrho$}, \mbox{\boldmath$\Delta$})=
V(\mbox{\boldmath$\varrho$}, \mbox{\boldmath$\Delta$})-
V(\mbox{\boldmath$\varrho$})
\nonumber \\
&&=\frac{2\varepsilon \varepsilon' n_a}{\omega m^4}
\int_{}^{}\left(1-\exp(i{\bf q} \mbox{\boldmath$\varrho$}) \right)
(d\sigma({\bf q}, \mbox{\boldmath$\Delta$})-
d\sigma({\bf q}, 0))d^2q,
\label{4.6}
\end{eqnarray}
where the cross section $d\sigma({\bf q}, \mbox{\boldmath$\Delta$})$ is
defined in Eq.(\ref{3.25}).
In the above formulas $\mbox{\boldmath$\varrho$}$ is the
two-dimensional space of the impact parameters measured in the
Compton wavelength $\lambda_c$ which is conjugate to the
two-dimensional space of the transverse momentum transfers ${\bf q}$
measured in the electron mass $m$.

\subsection{Amplitudes of coherent photon scattering under influence of
the multiple scattering}

When parameter $\Delta \ll m$ the main contribution into the integral
in (\ref{4.6}) gives the region $|{\bf q}| \leq \Delta/m \ll 1$. For 
$\varrho \leq 1$ one can expand
the integrand in (\ref{4.6}) in powers of
${\bf q} \mbox{\boldmath$\varrho$}$. Using Eq.(\ref{3.22}) we get
\begin{eqnarray}
&& \Delta V(\mbox{\boldmath$\varrho$}, \mbox{\boldmath$\Delta$})=
\frac{\varepsilon \varepsilon' n_a}{\omega m^4}
\int_{}^{} \frac{({\bf x}\mbox{\boldmath$\varrho$})^2}{x^2}(\chi'({\bf x}))^2
(\exp(-i\mbox{\boldmath$\delta$}{\bf x})-1)d^2x,
\nonumber \\
&& \chi({\bf x})=\int_{-\infty}^{\infty}\frac{Z\alpha}{r}
\exp\left(-\frac{r}{a_{s}} \right) = 2Z\alpha K_0\left(x
\frac{a_{s}}{\lambda_c}\right),\quad x=|{\bf x}|
\nonumber \\
&&\frac{a_{s}}{\lambda_c} = 111 Z^{-1/3}
=\frac{a_{s2}}{\lambda_c}\exp\left(f-\frac{1}{2} \right),\quad
\mbox{\boldmath$\delta$}=\frac{\mbox{\boldmath$\Delta$}}{m},
\label{4.7}
\end{eqnarray}
where $K_0(z)$ is the modified Bessel function.
Integrating $\Delta V$ over azimuthal angle of the vector ${\bf x}$
we obtain
\begin{eqnarray}
&& \Delta V (\mbox{\boldmath$\varrho$}, \mbox{\boldmath$\Delta$})=
-2Q \mbox{\boldmath$\varrho$}^2 \int_{0}^{\infty} K_1^2 (x)
\left(1-J_0(\beta x)+J_2(\beta x) \cos 2\varphi \right) xdx
\nonumber \\
&&=-2 Q \mbox{\boldmath$\varrho$}^2  \int_{0}^{\infty}
\left[ K_0^2(x) \cos 2\varphi+ K_1^2(x)\right]
\left(1- J_0(\beta x)\right) xdx
\nonumber \\
&&=Q \mbox{\boldmath$\varrho$}^2 \left[ F_1\left(\frac{\beta}{2} \right)
\cos 2\varphi+F_2\left(\frac{\beta}{2} \right) \right],\quad
\beta =\Delta a_{s}
\label{4.8}
\end{eqnarray}
where $\varphi$ is the angle between the vectors
$\mbox{\boldmath$\varrho$}$ and $\mbox{\boldmath$\Delta$}$,
$J_n(z)$ is the Bessel function, the functions $F_1(z)$ and $F_2(z)$
are defined in Eq.(\ref{3.43}). Some details of calculation are
given after Eq.(\ref{3.43}).

Summing $\Delta V(\mbox{\boldmath$\varrho$}, \mbox{\boldmath$\Delta$})$
(\ref{4.8}) and $V(\mbox{\boldmath$\varrho$})$ (\ref{4.5})
we get
\begin{equation}
V(\mbox{\boldmath$\varrho$}, \mbox{\boldmath$\Delta$})=
Q\mbox{\boldmath$\varrho$}^2
\left(L_1+\ln \frac{4}{\mbox{\boldmath$\varrho$}^2}-2C-
F_1\left(\frac{\beta}{2} \right)
\cos 2\varphi-F_2\left(\frac{\beta}{2} \right)\right)
\label{4.11}
\end{equation}

Starting from Eq.(\ref{4.2}) and advancing as in Sec.II (Eqs.(2.2)-(2.12))
of \cite{L0} we find
\begin{eqnarray}
&& M_{++}=\frac{2i\alpha m^2\omega}{n_a} \int_{}^{}
\frac{d\varepsilon}{\varepsilon \varepsilon'}
\left<0|s_1\left(G_{s}^{-1}-G_0^{-1} \right)+
s_2 {\bf p}\left(G_{s}^{-1}-G_0^{-1} \right){\bf p}|0\right>,
\nonumber \\
&& M_{+-}=\frac{2i\alpha m^2\omega}{n_a} \int_{}^{}
\frac{d\varepsilon}{\varepsilon \varepsilon'}
\left<0|s_3\left({\bf e}_{-}^{\ast}{\bf p}\right)
\left(G_{s}^{-1}-G_0^{-1} \right)\left({\bf e}_{+}{\bf p}\right)|0\right>,
\label{4.13}
\end{eqnarray}
where
\begin{equation}
s_1=1,~s_2=\frac{\varepsilon^2+\varepsilon'^2}{\omega^2},~
s_3=\frac{2\varepsilon \varepsilon'}{\omega^2},\quad
G_s={\cal H}+1,~G_0={\bf p}^2+1,
\label{4.14}
\end{equation}
${\cal H}$ is defined in Eq.(\ref{4.3}).
In derivation it is convenient to use properties of ${\bf e}_{\lambda}$
given in (\ref{3.33}). Note, that coefficients $s_1, s_2$ coincide
with the same coefficients in the theory of the LPM effect for
pair creation by a photon \cite{L5}. As in \cite{L1} (Eq.(2.17))
and \cite{L5} (Eq.(2.5))
we split the potential
$V(\mbox{\boldmath$\varrho$}, \mbox{\boldmath$\Delta$})$ as
\begin{eqnarray}
&& V(\mbox{\boldmath$\varrho$}, \mbox{\boldmath$\Delta$})
=V_c(\mbox{\boldmath$\varrho$}, \mbox{\boldmath$\Delta$})+
v(\mbox{\boldmath$\varrho$}, \mbox{\boldmath$\Delta$}),\quad
V_c(\mbox{\boldmath$\varrho$}, \mbox{\boldmath$\Delta$})=
q_p \mbox{\boldmath$\varrho$}^2,\quad q_p = QL_s,
\nonumber \\
&& L_s(\varrho_c)=\ln \frac{a_{s2}^2}{\lambda_c^2 \varrho_c^2}-
F_2\left(\frac{\beta}{2} \right),
\nonumber \\
&& v(\mbox{\boldmath$\varrho$}, \mbox{\boldmath$\Delta$})
=-\frac{q_p\mbox{\boldmath$\varrho$}^2}{L_s}
\left(2C+\ln \frac{\mbox{\boldmath$\varrho$}^2}{4\varrho_c^2}+
F_1\left(\frac{\beta}{2} \right)\cos 2\varphi \right)
\label{4.15}
\end{eqnarray}
In accordance with such division of the potential we will write the
photon scattering amplitudes in the form
\begin{equation}
M_{++}=M_{++}^c+M_{++}^{(1)},\quad M_{+-}=M_{+-}^c+M_{+-}^{(1)},
\label{4.16}
\end{equation}
where $M_{++}^c$ is the contribution of the potential
$V_c(\mbox{\boldmath$\varrho$}, \mbox{\boldmath$\Delta$})$
and $M_{++}^{(1)}$ is the first correction to the scattering amplitude
due to the potential $v(\mbox{\boldmath$\varrho$}, \mbox{\boldmath$\Delta$})$
(see \cite{L1}, Eqs.(2.28) and (2.33), and \cite{L5} Eqs.(2.10) and (2.15)).
Since dependence of the potential
$V(\mbox{\boldmath$\varrho$}, \mbox{\boldmath$\Delta$})$ on
$\mbox{\boldmath$\varrho$}$ is the same as in \cite{L5}
and the expression for $M_{++}$ (\ref{4.13}) formally coincides (up to
external factor) with
Eq.(2.2) in \cite{L5} the result for $M_{++}$ can be taken from
\cite{L5} (Eqs.(2.10) and (2.15)) with substitution $q \rightarrow q_p,~
L_c \rightarrow L_s$. The combination of photon polarization
vectors entering in the amplitude $M_{+-}^c$ (\ref{4.13})
gives in evaluation the expression
$\left({\bf e}_{-}^{\ast}\mbox{\boldmath$\nabla$}\right)
\left({\bf e}_{+}\mbox{\boldmath$\varrho$}\right)=0$.
Substituting this combination in calculation of the first correction
$M_{+-}^{(1)}$ (see Eqs.(2.31) and (2.32) of \cite{L1})
we obtain expression of type
$\left({\bf e}_{-}^{\ast}\mbox{\boldmath$\varrho$}\right)
\left({\bf e}_{+}\mbox{\boldmath$\varrho$}\right)=\mbox{\boldmath$\varrho$}^2
\exp(2i\varphi)$. So in integration over the azimuthal angle $\varphi$ only
the term with $\cos 2\varphi$ survives in formula for
$v(\mbox{\boldmath$\varrho$}, \mbox{\boldmath$\Delta$})$. As a result we get
for the main term
\begin{eqnarray}
&& M_{++}^c=\frac{\alpha m^2\omega}{2\pi n_a} \int_{}^{}
\frac{d\varepsilon}{\varepsilon \varepsilon'}
\Phi_s(\nu_s),\quad M_{+-}^c=0,
\nonumber \\
&&\displaystyle{\Phi_s(\nu_s)=\nu_s\int_{0}^{\infty} dt e^{-it}\left[s_1
\left(\frac{1}{\sinh z}-\frac{1}{z}\right)-i\nu_s s_2
\left( \frac{1}{\sinh^2z}- \frac{1}{z^2}\right) \right]}
\nonumber \\
&&=s_1\left(\ln p-\psi\left(p+\frac{1}{2}\right) \right)
+s_2\left(\psi (p) -\ln p+\frac{1}{2p}\right),
\label{4.17}
\end{eqnarray}
where $\nu_s=2\sqrt{iq_s}, z=\nu_s t,~p=i/(2\nu_s),~\psi(x)$
is the logarithmic derivative of the gamma function.

The first corrections to the amplitudes are defined by
\begin{eqnarray}
&& M_{++}^{(1)}=-\frac{\alpha m^2\omega}{4\pi n_a L_s} \int_{}^{}
\frac{d\varepsilon}{\varepsilon \varepsilon'}
F_s(\nu_s),\quad
M_{+-}^{(1)}=\frac{\alpha m^2\omega}{4\pi n_a L_s}
F_1\left(\frac{\beta}{2} \right) \int_{}^{}
\frac{d\varepsilon}{\varepsilon \varepsilon'}
F_3(\nu_s),
\nonumber \\
&& F_s(\nu_s)= \int_{0}^{\infty}\frac{dz e^{-it}}{\sinh^2z}
\left[s_1f_1(z)-2is_2f_2(z) \right],\quad
f_2(z) = \frac{\nu_s}{\sinh z}
\left(f_1(z)-\frac{g(z)}{2} \right),
\nonumber \\
&& f_1(z)=\left(\ln \varrho_c^2+\ln \frac{\nu}{i}
-\ln \sinh z-C\right)g(z) - 2\cosh z G(z),
\nonumber \\
&& F_3(\nu_s)=i \nu_s s_3\int_{0}^{\infty}\frac{dz e^{-it}}{\sinh^3z} g(z)
=\frac{s_3}{2}\left[1+\frac{1}{2p}-p~\zeta(2, p) \right]
\nonumber \\
&& g(z)=z\cosh z - \sinh z, \quad t=t_1+t_2,~z=\nu_s t,~p=\frac{i}{2\nu_s},
\nonumber \\
&& G(z)=\int_{0}^{z}(1-y\coth y)dy
\nonumber \\
&&\displaystyle{=z-\frac{z^2}{2}-\frac{\pi^2}{12}-
z\ln \left(1-e^{-2z} \right)
+\frac{1}{2}{\rm Li}_2 \left(e^{-2z} \right)},
\label{4.18}
\end{eqnarray}
here ${\rm Li}_2 \left(x \right)$ is the Euler dilogarithm,
$\zeta(s, a)$ is the generalized Riemann zeta function.
Use of the given representations of functions $F_3(\nu_s)$ and
$G(z)$ simplifies the numerical calculation.

In the region of the weak effect of scattering ($|\nu_s| \ll 1,~
\varrho_c=1$) the interval $z \ll 1$ contributes
into the integrals (\ref{4.17}), (\ref{4.18}). In this region
\begin{eqnarray}
&& \Phi_s(\nu) \simeq s_1\left(\frac{\nu^2}{6}+\frac{7\nu^4}{60}
+ \frac{31\nu^6}{126} \right)
+ s_2\left(\frac{\nu^2}{3}+\frac{2\nu^4}{15} + \frac{16\nu^6}{63} \right),
\nonumber \\
&& F_s(\nu) \simeq \frac{\nu^2}{9}\left(s_2-s_1\right),\quad
F_3(\nu) \simeq \frac{\nu^2}{3}\left(1+\frac{4}{5}\nu^2 + 
\frac{16}{7} \nu^4\right)s_3.
\label{4.19}
\end{eqnarray}
Substituting these expressions into (\ref{4.17}), (\ref{4.18})
and integrating over $\varepsilon$ we obtain
\begin{eqnarray}
\hspace{-10mm}&& M_{++}=M_{++}^c+M_{++}^{(1)}
=i\frac{14Z^2 \alpha^3 \omega}{9m^2}
\Bigg[L_{s1}\Bigg(1+i\frac{59\omega}{175 \omega_e}\frac{L_{s1}}{L_1}
\nonumber \\
\hspace{-10mm}&&-\frac{3312}{2401}\left(\frac{\omega}{\omega_e}
\frac{L_{s1}}{L_1}\right)^2
\Bigg)-\frac{1}{21} \Bigg],
\nonumber \\
\hspace{-10mm}&& M_{+-}=M_{+-}^{(1)}=i\frac{2Z^2 \alpha^3 \omega}{9m^2}
F_1\left(\frac{\beta}{2}\right)
\left(1+i\frac{16\omega}{25 \omega_e}\frac{L_{s1}}{L_1} -
\frac{384}{245}\left(\frac{\omega}{\omega_e}\frac{L_{s1}}{L_1} \right)^2
\right),
\label{4.20}
\end{eqnarray}
here
\begin{eqnarray}
&& L_{s1}=\ln \frac{a_{s2}^2}{\lambda_c^2}-F_2\left(\frac{\beta}{2}\right),
\quad \omega_e=\frac{m}{2\pi Z^2 \alpha^2\lambda^3n_a L_1},
\nonumber \\
&& L_{s1}-\frac{1}{21}= 2\left[\ln \frac{a_{s}}{\lambda_c}
-\frac{1}{2} \left(F_2\left(\frac{\Delta a_{s}}{2}\right)+1\right)
-f(Z \alpha)+\frac{41}{42} \right],
\label{4.21}
\end{eqnarray}
$L_1$ is defined in (\ref{4.5}).
The characteristic energy $\omega_e$ encountered in analysis of influence
of the multiple scattering on the probability of pair 
photoproduction \cite{L5}, in gold $\omega_e$=10.5~TeV.
The amplitudes (\ref{4.20}) coincide with the formulas (\ref{3.47}) and
(\ref{3.55}) if we neglect the terms $\propto \omega/\omega_e$
and $(\omega/\omega_e)^2$.
The terms $\propto \omega/\omega_e$ define the real part of the 
scattering amplitudes while the terms $\propto (\omega/\omega_e)^2$
are the corrections to the imaginary part.

If the parameter $|\nu_s| > 1$ then the value of $\varrho_c$
is defined by the equations (see (\ref{4.15}))
\begin{eqnarray}
&& 4Q\varrho_c^4L_s(\varrho_c)=1,\quad
L_{s} \simeq L_{s3} + \frac{1}{2} \ln \frac{4\varepsilon \varepsilon'}
{\omega^2},\quad L_{s3}=L_{s2}+\frac{1}{2} \ln L_{s2},
\nonumber \\
&& L_{s2}=L_{s1}+\frac{1}{2}\ln \frac{\omega}{\omega_e}=
2\ln(a_{s}\Delta_s)+1-F_2\left(\frac{\Delta a_{s}}{2}\right)-2 f(Z\alpha),
\nonumber \\
&& \Delta_s^4=2\pi Z^2 \alpha^2 \omega n_a,\quad
\nu_s^2=i\frac{\omega}{\omega_e}\frac{4\varepsilon \varepsilon'}{\omega^2}
\left(L_{s3}+ \frac{1}{2} \ln \frac{4\varepsilon \varepsilon'}
{\omega^2}\right).
\label{4.22}
\end{eqnarray}

We consider now the region of the strong effect of scattering
($|\nu_s| \gg 1$). We restrict to the main terms of the decomposition only,
then we can substitute the exponential $\exp (-it)$ for 1 in integrand
in Eqs.(\ref{4.17}),(\ref{4.18}). Performing the integration we find
\begin{equation}
\Phi_s(\nu_s) \simeq is_2\nu_s,\quad F_s(\nu_s) \simeq
-is_2\nu_s\left(\ln 2 -C -i\frac{\pi}{4}\right),\quad F_3(\nu_s) \simeq
\frac{i}{2}s_3 \nu_s.
\label{4.23}
\end{equation}
Note, that the next terms of the decomposition can be obtained using the
results of Appendix A of \cite{L5}.
Substituting these expressions into Eqs.(\ref{4.16})-(\ref{4.18})
we obtain
\begin{eqnarray}
&& M_{++} \simeq (i-1)\frac{3\alpha m^2}{4\sqrt{2}n_a}
\sqrt{\frac{\omega}{\omega_e}\frac{L_{s3}}{L_1}}\left[
1-\frac{1}{4L_{s3}}\left(2C+\frac{1}{3}+i\frac{\pi}{2} \right) \right]
\nonumber \\
&& =(i-1)\frac{3\pi Z^2 \alpha^3 \omega}{2\sqrt{2}\Delta_s^2}
\sqrt{L_{s3}}\left[
1-\frac{1}{4L_{s3}}\left(2C+\frac{1}{3}+i\frac{\pi}{2} \right) \right],
\nonumber \\
&& M_{+-} \simeq (i-1)\frac{\pi Z^2 \alpha^3 \omega}{8\sqrt{2}\Delta_s^2}
\frac{1}{\sqrt{L_{s3}}}F_1\left(\frac{\beta}{2}\right).
\label{4.24}
\end{eqnarray}
It is seen from these equations that in the case of strong scattering
the real and imaginary parts of the amplitudes are equal (if we neglect
the term $\propto 1/L_{s3}$ in $M_{++}$). Moreover the amplitudes
(\ref{4.24}) don't depend on the electron mass $m$. In place of it we have
the value $\Delta_s$. Notice that the asymptotic expansions (\ref{4.24})
are valid for $\Delta \ll \Delta_s~(\Delta_s > m)$. In the interval
$\Delta_s \gg \Delta \gg a_{s}^{-1}$ we have
\begin{equation}
F_2 \simeq 2\ln(\Delta a_{s})-1,\quad F_1=1,\quad
L_{s2} = 2\left[\ln \frac{\Delta_s}{\Delta}+1-f(Z \alpha) \right].
\label{4.25}
\end{equation}

\section{Conclusion}
\setcounter{equation}{0}

In this paper we considered the influence of a medium on
the process of the coherent photon scattering
illustrated in Fig.1 and Fig.2,
where Im~$M_{++}$ and Re~$M_{++}$ as well as Im~$M_{+-}$ and Re~$M_{+-}$ are
given as a function of photon energy $\omega$ in gold.
This influence is due to the multiple
scattering of electron and positron of the virtual pair on the formation
length of the process (see Eqs.(\ref{1.1}) and (\ref{1.2}))
\begin{equation}
l_f=\frac{\omega}{2(q_s^2+m^2+\Delta^2)}.
\label{5.1a}
\end{equation}
In the region $\Delta^2 \ll q_s^2+m^2$ this formation length
is independent of $\Delta$ and it value is coincide practically with
the formation length of pair creation by a photon $l_c$ \cite{L5}.
There is some difference connected with the logarithmic dependence of
$\nu_s^2$ value on $\Delta^2$:
\begin{equation}
|\nu_s^2|=\frac{q_s^2}{m^2}=\frac{4\pi Z^2 \alpha^2}{m^2}n_a l_f
\int_{q_{min}^2}^{q_{max}^2}\frac{dq^2}{q^2}
\label{5.1b}
\end{equation}
where $q_{max}^2=m^2+q_s^2$ and $q_{min}^2=\Delta^2+a^{-2}$, $a$ is
the screening radius of atom (\ref{1.2}). This defines the weak
(logarithmic) dependence of Im$M_{++}$ on $\Delta$ in the region
$\Delta < \sqrt{q_s^2+m^2}$.
This can be seen in Fig.1 where the curves 1 and 3 represent behavior
of Im~$M_{++}$ for $\Delta$=0.4435 $m$ and $\Delta$=0.0387 $m$ respectively.
For lower value of $\Delta$ the minimal momentum transfer $q_{min}$
(\ref{1.2}) diminishes thereby the interval of contributing
the multiple scattering angles increases, so the multiple scattering
affects the photon scattering amplitude at a lower energy (and smaller
formation length). Because of this the curve 3 is shifted to the
left respect the curve 1. Note that the curve 3 ($\Delta^{-1} \sim a_{s2}$)
is very similar to the curve 2 in Fig.2 of \cite{L5} which represents the
behavior of the probability of pair photoproduction in gold vs photon
energy.

The new property of influence of a medium
is the appearance of the real part of the 
coherent photon scattering amplitudes at high energy $\omega$.
In the region $\omega \ll \omega_e$  the value of Re$M$ is small
accordingly (\ref{4.20}). In the asymptotic region
$\omega \gg \omega_e$ we have -Re~$M =$ Im~$M$ according to
(\ref{4.24}). This property is seen clearly in Figs.1,2.
So the value of -Re~$M$ is small at low and very high energies of photon.
At intermediate energies the value of -Re~$M$ have the maximum
at $\omega \simeq$220~TeV for $\Delta=$0.4435 $m$ and
at $\omega \simeq$80~TeV for $\Delta=$0.0387 $m$.
In Fig.2 the same curves are shown for amplitude $M_{+-}$.
These curves are very similar to curves in Fig.1.
The curves in both figures are normalized to imaginary part of the
corresponding amplitude in the absence of the multiple scattering.
The ratio these imaginary parts one can find from Eqs.(\ref{4.20}),
(\ref{4.21}):
\begin{equation}
r=\frac{{\rm Im~}M_{+-}}{{\rm Im~}M_{++}}=
\frac{1}{14}\displaystyle{F_1\left(\frac{\Delta a_{s}}{2} \right)
\left[\ln \frac{a_{s}}{\lambda_c}-\frac{1}{2}\left[
F_2\left(\frac{\Delta a_{s}}{2} \right)+1 \right]-f(Z\alpha)+1 \right]^{-1}}.
\label{5.1}
\end{equation}
This ratio is $r=0.04435$ for $\Delta$=0.4435 $m$ and
$r=0.003018$ for $\Delta$=0.0387 $m$.

In numerical calculation of amplitude $M_{++}$ we neglect corrections
of the order $1/L_s$. These corrections are quite small, e.g. for
$\Delta$=0.0387 $m$ they are of the order of a few percent.

We estimate now the integral (over $\mbox{\boldmath$\Delta$}$)
cross section of the coherent photon scattering at $\omega \gg \omega_e$.
From previous section (see Eqs.(\ref{4.24})-(\ref{4.25})) we have for
module squared of the amplitude at $\Delta < \Delta_s$
(see also Introduction)
\begin{equation}
|M|^2 \simeq \frac{9\pi^2Z^4\alpha^6\omega^2}{\Delta_s^4}
\left(\ln \frac{\Delta_s}{\Delta}+1\right),
\label{5.2}
\end{equation}
where value $\Delta_s$ is defined in (\ref{4.22}).
It is seen from this formula that in the considered region
the differential probability of the coherent photon scattering depends weakly
(logarithmically) on medium density $n_a$.
\begin{equation}
dW_{ph}(\Delta < \Delta_s)=\frac{1}{16\pi^2}|M|^2 n_a d\Omega
=\frac{|M|^2n_a}{16\pi^2 \omega^2}d\mbox{\boldmath$\Delta$}.
\label{5.3}
\end{equation}

In the interval $\Delta \gg \Delta_s$ the influence of the multiple scattering
on the coherent scattering process is rather weak because the probability
of transfer to a medium of the momentum $\geq |\mbox{\boldmath$\Delta$}|$
on the formation length $l_f=\omega/(2\Delta^2)$ appears to be small:
\begin{equation}
W_s(\Delta)=\frac{4\pi Z^2\alpha^2}{\Delta^2}n_a\frac{\omega}{2\Delta^2}
=\frac{\Delta_s^4}{\Delta^4} \ll 1.
\label{5.4}
\end{equation}
In this interval the amplitudes of the coherent photon scattering
behave as $1/\Delta^2$ and it doesn't contribute to the
integral cross section. So the interval $\Delta \leq \Delta_s$
contributes only. Taking into account Eqs.(\ref{5.2}),(\ref{5.3})
we have for estimate of the integral cross section of 
the coherent photon scattering
\begin{equation}
\sigma_{ph}(\omega \gg \omega_e) \sim \frac{\Delta_s^2}{16\pi \omega^2}
|M(\Delta \sim \Delta_s)|^2 \sim \frac{Z^4\alpha^6}{\Delta_s^2}
=\frac{Z^3\alpha^5}{\sqrt{2\pi \omega n_a}}.
\label{5.5}
\end{equation}

{\bf Acknowledgments}
\vspace{0.2cm}
This work
was supported in part by the Russian Fund of Basic Research under Grant
00-02-18007.

\newpage

{\bf Figure captions}

\vspace{15mm}
\begin{itemize}

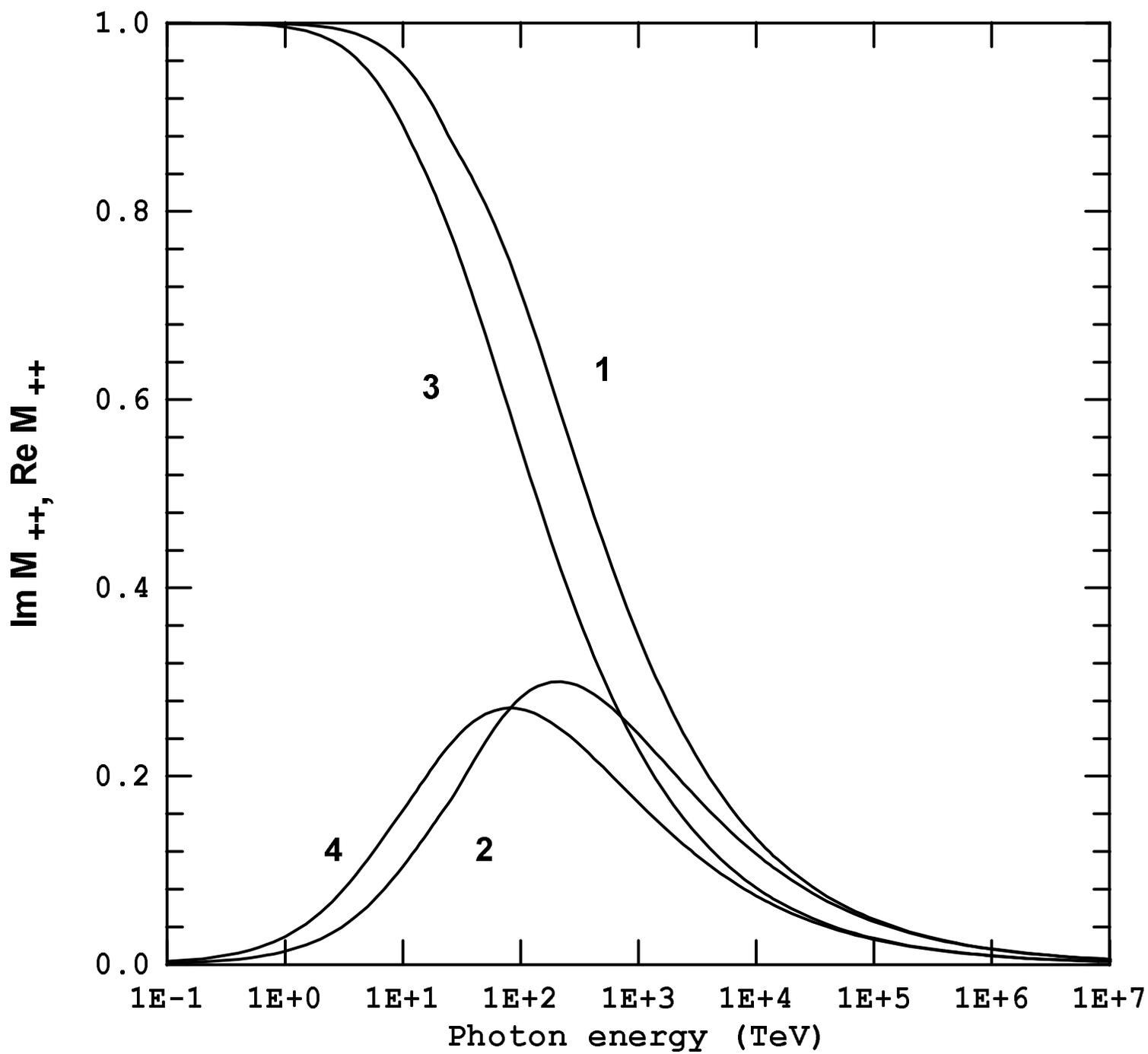
\item {\bf Fig.1} The amplitude $M_{++}$ of the coherent photon scattering
in gold under influence of the multiple scattering at the different
momentum transfer to the photon $\Delta$ in terms of the amplitude
Im$M_{++}$ (\ref{3.55}) calculated for the screened Coulomb
potential.
\begin{itemize}
\item Curve 1 is Im$M_{++}$ for $\Delta=$0.4435 $m$.
\item Curve 2 is Re$M_{++}$ for $\Delta=$0.4435 $m$.
\item Curve 3 is Im$M_{++}$ for $\Delta=$0.0387 $m$.
\item Curve 4 is Re$M_{++}$ for $\Delta=$0.0387 $m$.
\end{itemize}

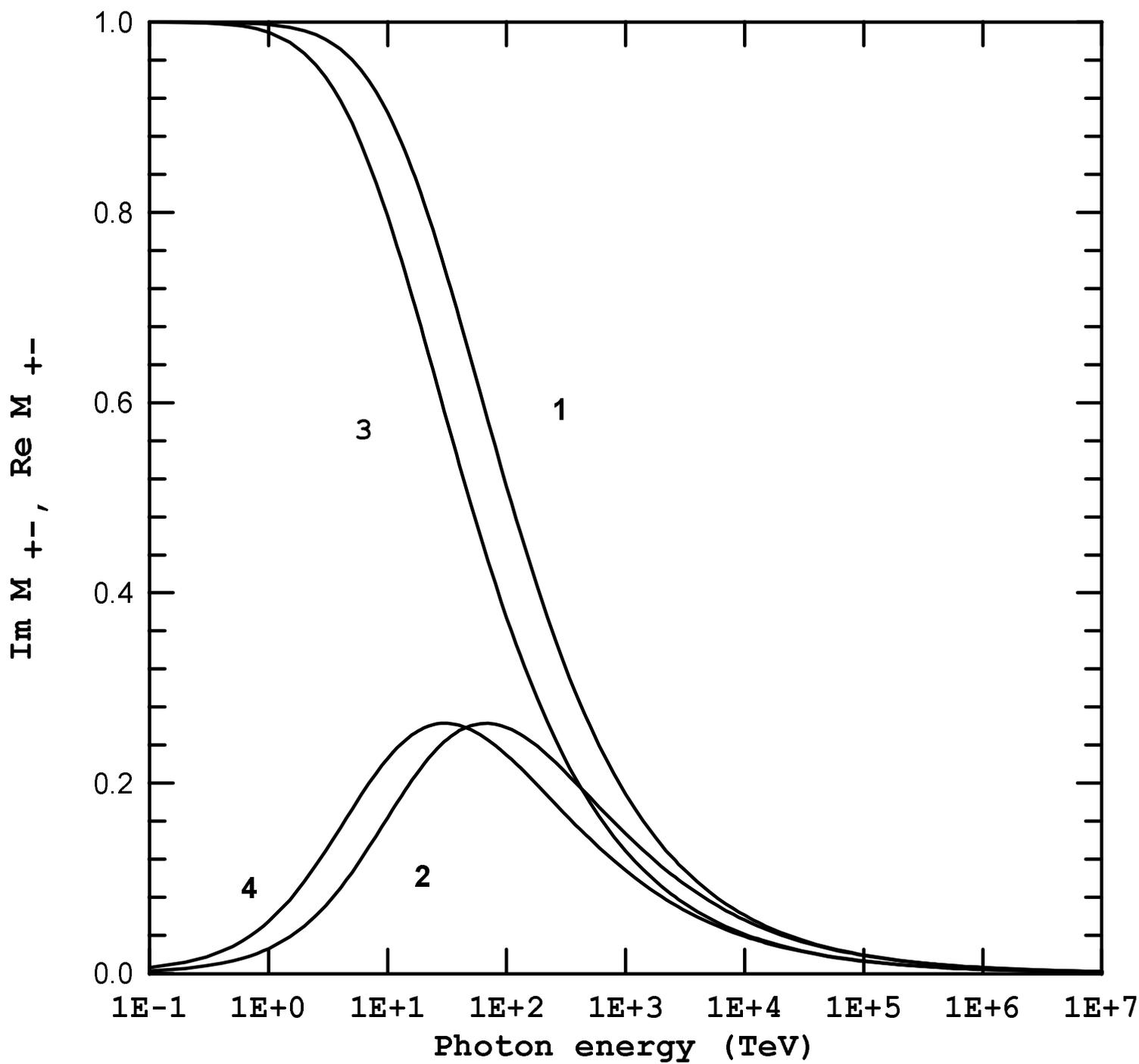
\item {\bf Fig.2} The same as in Fig.1 but for
the amplitude $M_{+-}$ in terms of the amplitude
Im$M_{+-}$ (\ref{3.55}) calculated for the screened Coulomb
potential.

\end{itemize}

\end{document}